\documentclass[a4paper,11pt]{article}
\usepackage{amsmath}
\usepackage{amsfonts}
\usepackage{amssymb}
\usepackage{bm}
\usepackage[utf8x]{inputenc}
\usepackage{ae}
\usepackage[T1]{fontenc}
\usepackage[english]{babel}
\usepackage{graphicx}
\usepackage{makecell}
\usepackage{placeins}
\usepackage{tabularx}
\usepackage{tabulary}
\usepackage{setspace}
\usepackage{tablefootnote}
\usepackage[flushmargin]{footmisc}
\usepackage[comma]{natbib}
\usepackage{array}
\usepackage{float}
\usepackage{caption}
\usepackage{tikz}
\usetikzlibrary{shapes.geometric}
\usepackage[a4paper,left=2.5cm,right=2.5cm,top=3cm,bottom=3cm]{geometry}
\usepackage[colorinlistoftodos]{todonotes}
\usepackage{pdfpages}
\usepackage{longtable}
\usepackage{adjustbox}
\usepackage{booktabs}
\usepackage{multirow}
\usepackage{lscape}
\usepackage[colorlinks=true, allcolors=blue]{hyperref}

\usepackage[flushleft]{threeparttable}
\usepackage{inputenc}



\begin{document} \doublespacing \pagestyle{plain}
	
	\def\ci{\perp\!\!\!\perp}
	\begin{center}
		
		{\LARGE The Effect of the Gotthard Base Tunnel on Road Traffic: A Synthetic Control Approach
		}

		{\large \vspace{0.8cm}}
		
		{\large Hannes Wallimann, Widar von Arx, and Ann Hesse}\medskip

		{\small {University of Applied Sciences and Arts Lucerne, Competence Center for Mobility} \bigskip }
		
		{\large \vspace{0.8cm}}
		
		{\small {Version June 2025} \bigskip }

	\end{center}
	
	\smallskip

	\small \noindent \textbf{Abstract:} {The opening of the Gotthard Base Tunnel in 2017, the longest railway tunnel in the world, marked a milestone in Swiss transport policy. The tunnel, a part of the New Rail Link through the Alps, serves as a key instrument of the so-called "modal shift policy," which aims to transfer transalpine freight traffic from road to rail. The reduction in travel time by train between northern and southern Switzerland raised expectations that a substantial share of tourist-oriented passenger traffic would also shift from car to rail. In this paper, we conduct a causal analysis of the impact of the Gotthard Base Tunnel's opening at the end of 2016 on the number of cars using the parallel Gotthard motorway section in the subsequent years. To this end, we apply the synthetic control and the synthetic difference-in-differences methods to construct a synthetic Gotthard motorway section based on a weighted combination of other alpine road crossings (a so-called donor pool) that did not experience the construction of a competing rail infrastructure. Our results reveal only a modest but statistically significant decline in the number of cars between the actual and the synthetic Gotthard motorway in the short run. Given the consistently strong and increasing demand for the new rail connection through the Gotthard Base Tunnel, we infer a substantial induced short-run demand effect resulting from the rail travel time savings.}
	
	{\small \smallskip }
	{\small \smallskip }
	{\small \smallskip }
	
	
	{\small \smallskip }
	{\small \smallskip }
	{\small \smallskip }
	
	{\small \noindent \textbf{Acknowledgments:} We are indebted to Antonin Danalet for helpful comments. Moreover, Zeliya Schär provided excellent research assistance. }
	
	\bigskip
	\bigskip
	\bigskip
	\bigskip
	
	{\small {\scriptsize 
\begin{spacing}{1.5}\noindent  
\textbf{Addresses for correspondence:} Hannes Wallimann, University of Applied Sciences and Arts Lucerne, Rösslimatte 48, 6002 Lucerne, \href{mailto:hannes.wallimann@hslu.ch}{hannes.wallimann@hslu.ch}.
\end{spacing}
			
		}\thispagestyle{empty}\pagebreak  }

	{\small \renewcommand{\thefootnote}{\arabic{footnote}} %
		\setcounter{footnote}{0}  \pagebreak \setcounter{footnote}{0} \pagebreak %
		\setcounter{page}{1} }
	
\section{Introduction}\label{introduction}

The rationale for the construction of an efficient railway infrastructure differs across countries. For Switzerland, investments in a flat, energy-efficient, and high-performance railway infrastructure are conceived as an important contribution aimed at limiting the negative environmental impact on the sensitive alpine landscape of road transport \citep{albalate2012high}. However, robust ex-post analyses of the impact of new rail infrastructure on induced demand and modal shift from private car to rail remain scarce. To the best of our knowledge, only the study of \citet{borsati2020modal} systematically examines the causal impact of high-performance railway infrastructure on motorway, concluding that neither the opening of a high-speed rail (HSR) nor the opening of on-track competition led to a modal shift from motorway to HSR services. Moreover, the review of \citet{givoni2013review} on HSR indicates that the demand for HSR several years after inauguration is about ten to twenty percent induced demand. The rest is attributed to mode substitution, whereas, in most cases, most HSR passengers have traveled with conventional rail before. In contrast, the substitution from aircraft, car, and coach is usually modest, although the effects vary across contexts and studies \citep{givoni2013review}.

These findings underscore the need for a new, rigorous and empirically driven evaluation of infrastructure policies. To fill this research gap, we focus on the Gotthard Base Tunnel, a part of the New Rail Link through the Alps. The construction of the Gotthard Base Tunnel took 17 years and was completed in December 2016, at a total cost of 12.2 billion Swiss francs. In our study, we answer the research question of whether the travel time savings due to the new Gotthard Base Tunnel in the short run reduced the number of light vehicles on the parallel motorway section. Therefore, knowing the average car occupancy rate in leisure travel, we can also approximate whether the new connection induced an increase in passengers. 

The case of the Gotthard Base Tunnel is interesting for several reasons. While most research has focused on inter-modal competition between HSR and air services, especially on long point-to-point links between cities, our case of the Gotthard Base Tunnel differs in relevant aspects. The section with the Gotthard Base Tunnel from Art-Goldau to Bellinzona is relatively short, with a travel time of 58 minutes, which means that the private car is a relevant competitor. Such studies are relatively scarce \citep[see for an example the study of ][]{borsati2020modal}. Most researched HSR connections have a length of several hundred kilometers \citep{cascetta2011analysis,cheng2010high,givoni2013review}. Second, Switzerland has no HSR trains, only an intercity network. Third, the Swiss rail system is “open”, which means that customers can board a train at any time without a reservation. Fourth, there are no special prices for the Gotthard Base Tunnel connection. Together with the fact that Switzerland has a dense suburban public transport service to every municipality, this eliminates the problem of many of the HSR examples studied being difficult for customers to access or having no connection to the countryside at the destination \citep{sanchez2012accessibility}. Another distinguishing feature is the touristic use of the route, including “transit traffic”; for example, tourists from Germany travel to Italy through the Gotthard axis, crossing the Swiss Alps. A recent study has shown that 97 percent of these people travel by private car and only 3 percent by train \citep{ohnmacht2024definition}. 

In our comparative case study, we apply the synthetic control \citep{abadie2010synthetic} and the synthetic difference in differences \citep{arkhangelsky2021synthetic} methods to construct a synthetic Gotthard motorway section---a counterfactual that estimates the number of vehicles that would have used the Gotthard alpine crossing in the absence of the tunnel. The synthetic Gotthard motorway section is a weighted average of other alpine crossings in Switzerland and neighboring countries, serving as the donor pool. Our results indicate that, on average, the number of light vehicles on the Gotthard motorway section decreased by 135 or 152 per day, depending on the method, during the months of April to October. This---using the numbers of leisure travelers---corresponds to around 300 fewer people traveling by car through the Gotthard every day. These effects correspond to a just under 1\% reduction compared to the pre-treatment average observed in the years prior to the opening of the Gotthard Base Tunnel. However, despite the modest size of the effect, the 95\% bootstrap confidence intervals amount to [-269; -127] and [-291; -84], with all values below zero. Sensitivity analyses, e.g., when controlling for potential spillover effects, confirm the robustness of this small but statistically significant effect on the parallel motorway section. Moreover, when comparing the Gotthard motorway section with other Swiss motorway sections prone to leisure traffic---instead of other alpine crossings---, the negative effect on daily vehicles increases to 617. However, this effect is not statistically significant. When comparing these numbers with the simultaneous daily increase of approximately 2,000 rail passengers (e.g., from 2016 to 2017), we conclude that induced traffic is the main part of the additional passengers on the Gotthard rail route. 

Our paper proceeds as follows. In Section \ref{Literature}, we discuss the related literature for our case study. Section \ref{Background} introduces the Gotthard Base Tunnel in greater detail. In Section \ref{Methods}, we present our research design, outlining both the identification strategy and the econometric method used to estimate the causal effect of interest. Section \ref{Data} shows the data and first descriptive results. In Section \ref{Results}, we present the main results and those of the sensitivity analyses. Section \ref{Discussion} discusses the results and concludes. 

\section{Literature review}\label{Literature}

In the context of tourism, for travel distances up to 400 km, the main distinction in travel mode choice is between the car and public transport, while for journeys exceeding 400 km, air transport becomes the dominant mode \citep{thrane2015examining}. Regarding mode choice between car and public transport---as with mode choice between other transport modes---one can differentiate between push measures that discourage car use and pull measures that encourage public transport use \citep[see, e.g., ][]{zarabi2024enhancing}. However, a research gap exists concerning hard pull measures that intervene at the level of physical infrastructure. It is well known that demand, measured, for example, by the number of travelers, increases when infrastructure is improved, as shown by, e.g., \citet{pagliara2015high}. Also, several studies show that demand for public transport increases due to large-scale rail investments, e.g., the high-speed rail (HSR) in Europe increased inter-city traffic between major metropolitan areas in North-west Europe \citep[see, e.g., ][]{vickerman2015high}. According to \citet{givoni2013review}, between 75\% and 90\% of new HSR demand is drawn from other transport modes. However, the extent and source of modal shift vary significantly across routes, depending on factors such as route distance, number of stops, travel time, and the speed of the HSR service. Depending on these characteristics, the shift may primarily occur from air travel, conventional rail, coach, or private car. HSR services tend to achieve the highest market shares when in-vehicle travel time ranges between 1 and 3.5 hours \citep{givoni2013review}. 

However, the literature on causal ex-post analyses of the impact of new rail infrastructure on modal shift and induced demand from private cars to rail remains scarce. An exception is the study of \citet{borsati2020modal} that systematically examines the causal impact of high-performance railway infrastructure on motorways. \citet{borsati2020modal} conclude that neither the opening of a HSR nor the opening of on-track competition led to a modal shift from motorway to HSR services. Our study contributes to this literature by analyzing the Gotthard Base Tunnel, a tourist route that includes transit traffic.

In their review, \citet{givoni2013review} emphasizes the difficulty of identifying a definitive set of factors that explain modal substitution and induced demand following the introduction of high-speed rail. The considerable variation between HSR corridors suggests that local contextual factors, such as the characteristics of the connection or spatial and economic conditions, play an important role. Nonetheless, certain determinants are consistently recognized in literature. The most influential factor is travel time, which is directly linked to the operational speed differentials among transport modes \citep{cascetta2011analysis}. According to a survey by \citet{coto2007effects}, other important factors include comfort, service frequency, and the novelty of the HSR experience, whereas ticket prices were mentioned less frequently. A critical barrier to HSR adoption is the access and egress time to and from the railway station \citep{chang2008accessibility}. Private cars become more cost-effective when travel costs are shared among multiple passengers. Additionally, they offer greater flexibility and convenience, particularly in terms of luggage transport \citep{cascetta2011analysis}. Finally, environmental attitudes may also influence mode choice, with a preference for rail over car or air travel among passengers with a stronger “green” orientation. However, empirical research shows that this connection between attitude and action often does not exist \citep{alcock2017green}.

Regarding study design, our study adds to a novel branch of literature that employs the synthetic control method to measure the effect of transport infrastructure investment as a policy intervention on selected economic variables relevant to the geographical space of interest. Recent work includes the study by \citet{doerr2020new}, which applies the synthetic control method to assess the effect of a new regional commercial airport in the German state of Bavaria on tourism. The results indicate that the new transportation infrastructure increased tourism---measured by arrivals--- and promoted regional economic development in the Allgäu region over the period 2008 to 2016. Moreover, \citet{li2021estimating} use the synthetic control method to estimate the causal effects of the Belt and Road Initiative (BRI) on, e.g., local employment in the treatment regions Duisburg (Germany) and Piraeus (Greece). The results indicate that quantitative evidence of the BRI project on employment is small in the short run while (maritime) freight volumes increase. Finally, \citet{lahura2023effect} investigate the effect of the construction of Peru's first cable car system on tourism demand---measured by the number of visitors. The results indicate that the number of visitors more than doubled due to the transport infrastructure investments of the Kuelap Archaeological Complex.

\section{Background}\label{Background}

Switzerland, located in central Europe, comprises 26 federal states, also known as cantons. The country acknowledges four official national languages: German, French, Italian, and Romansh. The Canton of Ticino is unique as the only canton where Italian is the exclusive language. By the end of the third quarter of 2024, Ticino had about 359,000 inhabitants, accounting for 4\% of the Swiss population and constituting the majority of Italian-speaking Switzerland.\footnote{See \url{https://www.bfs.admin.ch/bfs/de/home/statistiken/bevoelkerung.assetdetail.33248190.html} (accessed on December 4, 2024).} It is situated on the southern side of the Alps and is predominantly surrounded by Italy.

\begin{figure}[H]
	\centering \caption{\label{InfrastructureInvestment} Spending on new transport construction and the improvement of the existing network in Switzerland (Source: OECD)}	\includegraphics[scale=.25]{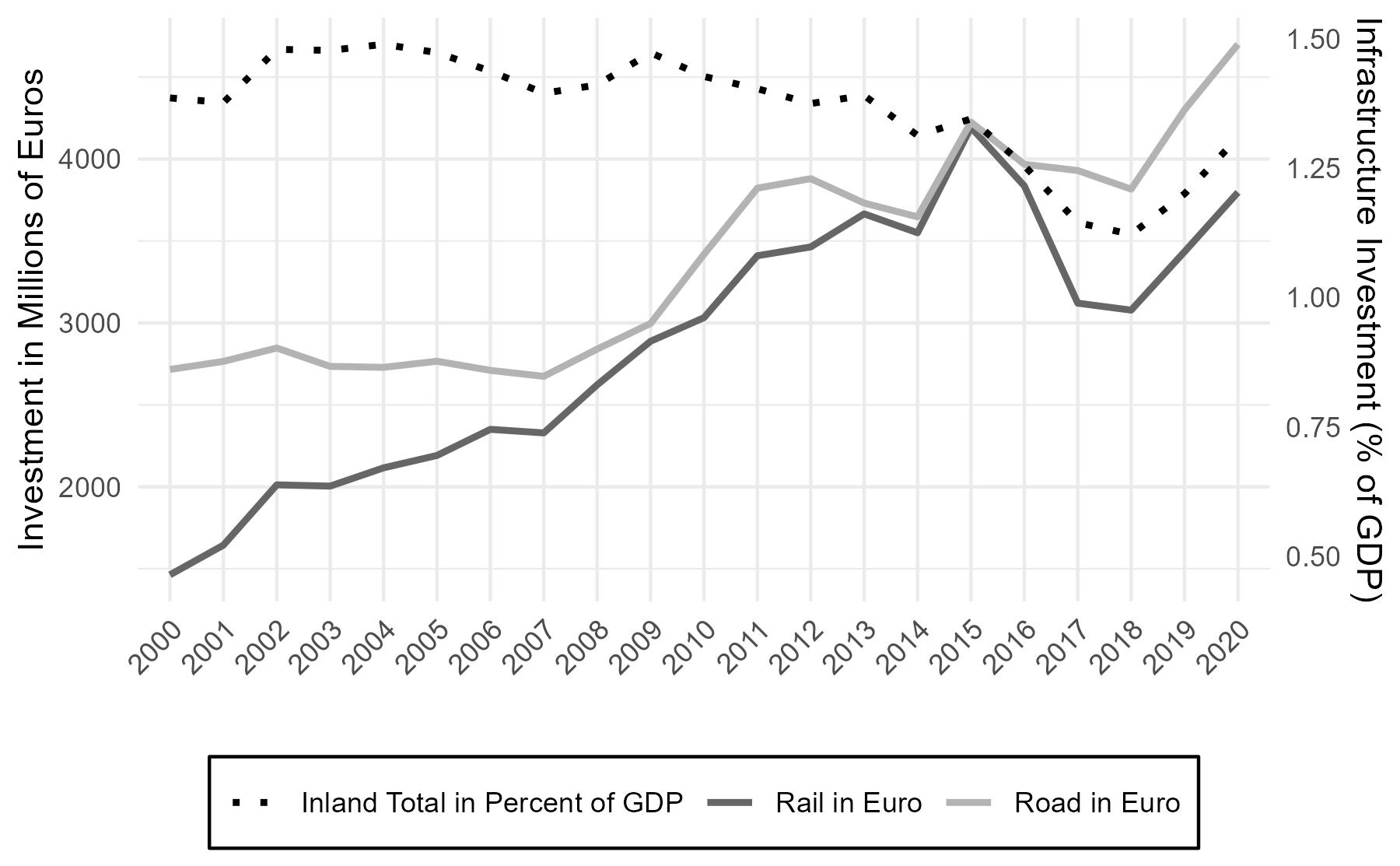}
\end{figure}

Switzerland boasts a good transport infrastructure that encompasses 85,009 km of road (2024), 5,317 km of railway track (2020), and 14 national and regional airports (2023).\footnote{See \url{https://www.bfs.admin.ch/bfs/en/home/statistics/mobility-transport/transport-infrastructure-vehicles.html} (accessed on December 10, 2024).} According to the OECD, in 2020, Switzerland invested about 3,667 million Euros in rail and 4,700 million Euros in road infrastructure, including new transport construction and improvement of existing networks.\footnote{See \url{https://www.oecd.org/en/data/indicators/infrastructure-investment.html?} (accessed on December 12, 2024). Note that the figures for the OECD's road infrastructure do not include operational expenditures, such as costs for routine maintenance.} Figure \ref{InfrastructureInvestment} displays trends in infrastructure rail and road investments in Switzerland from 2000 to 2020, both solid lines. While investments in rail and road (shown in millions of Euros) generally increased over time, the total infrastructure investments---including air and inland waterways infrastructure---as a percentage of GDP, dotted line, show fluctuations with a decline after 2009 and a subsequent recovery. 

The Swiss political system operates as a direct democracy. On September 27, 1992, Swiss voters approved the New Rail Link through the Alps\footnote{In German: Neue Eisenbahn-Alpentransversale (NEAT).} (NRLA) initiative, with 64 percent voting in favor. The NRLA represents Switzerland's largest construction project in its history.\footnote{See \url{https://www.eda.admin.ch/aboutswitzerland/en/home/swiss-stories/bauwerk-im-dienste-europas.html} (accessed on December 4, 2024).} It includes three tunnels: the Lötschberg Base Tunnel (34.6 km), the Gotthard Base Tunnel (57.1 km)---the longest railway tunnel in the world---, and the Ceneri Base Tunnel (15.4 km). Switzerland invested approximately 22.8 billion Swiss Francs in the construction of the NRLA, nearly 3.5 \% of its Gross Domestic Product (GDP). The Lötschberg Base Tunnel went into operation in 2007.\footnote{The number of public transport passengers increased by 59\% within one year \citep[see, e.g., ][]{danalet2025presentation}.}

The NRLA facilitates the operation of longer and heavier trains, allowing more freight trains to traverse the Alps with fewer locomotives and in less time. This enhancement not only accelerates passenger travel times but also offers a wider range of connections within Switzerland and across Europe. Moreover, the Gotthard route's rail transport capacity has greatly increased: up to 260 freight trains and 65 passenger trains can now travel daily at speeds up to 250 km/h (whereas in reality, due to efficiency reasons, they usually travel daily at speeds up to 200 km/h). With the opening of the Gotthard Base Tunnel, there was initially no increase in the number of trains. An (major) expansion of rail services only occurred following the commissioning of the Ceneri Base Tunnel. 

The Gotthard Base Tunnel's construction cost 12.2 billion Swiss Francs and took 17 years. It connects the Canton of Ticino with the German-speaking part of Switzerland, bringing northern and southern Europe closer together. Scheduled passenger services through the Gotthard Base Tunnel commenced in December 2016. Depending on the destination, passengers experienced travel time savings of approximately 20 to 40 minutes (see Table \ref{tab:travel_time_gbt} in Appendix \ref{Appendix_Tables}). The previous “mountain route” over the Gotthard, i.e., the “normal” train, was not neglected after the opening of the tunnel; on the contrary, it was upgraded with an attractive tourist train (with plenty of space for bicycles). Therefore, the frequent observation that an HSR connection leads to the decline of the existing service \citep[see, e.g., ][]{givoni2013review} is not true in this example. In August 2023, a wheel fracture occurred on a freight train wagon in the Gotthard Base Tunnel. This incident, which resulted in no injuries, led to a year-long disruption due to the challenging cleanup and repair efforts. During this period, the travel time for passenger trains increased by one hour. According to Ticino Turismo, the number of daily tourists from German-speaking Switzerland decreased by 15 percent during this time.\footnote{See \url{https://www.srf.ch/news/schweiz/gotthardtunnel-nach-unterbruch-gotthard-ist-offen-was-das-fuer-pendler-und-wirtschaft-bedeutet} (accessed on December 4, 2024).} 

At the end of 2020, the Ceneri Base Tunnel was opened to traffic. Together with the Gotthard Base Tunnel, it reduced travel time by one hour between Zurich and Milan by public transport. In addition, the Ceneri Base Tunnel enhanced the attractiveness of the regional rail network in the canton of Ticino. However, since travel behavior---particularly for leisure purposes---was affected by the COVID-19 pandemic in 2020, the Ceneri Base Tunnel is not considered in the present comparative case study on the causal effect of the Gotthard Base Tunnel on the parallel motorway section.

\begin{figure}[H]
	\centering \caption{\label{Passengers_rail} Development of Daily Rail Traffic Volumes: Gotthard vs. overall Long-Distance Travel Swiss Federal Railways (Indexed, 2013 = 100; figures were provided by the Swiss Federal Railways)}\includegraphics[scale=1]{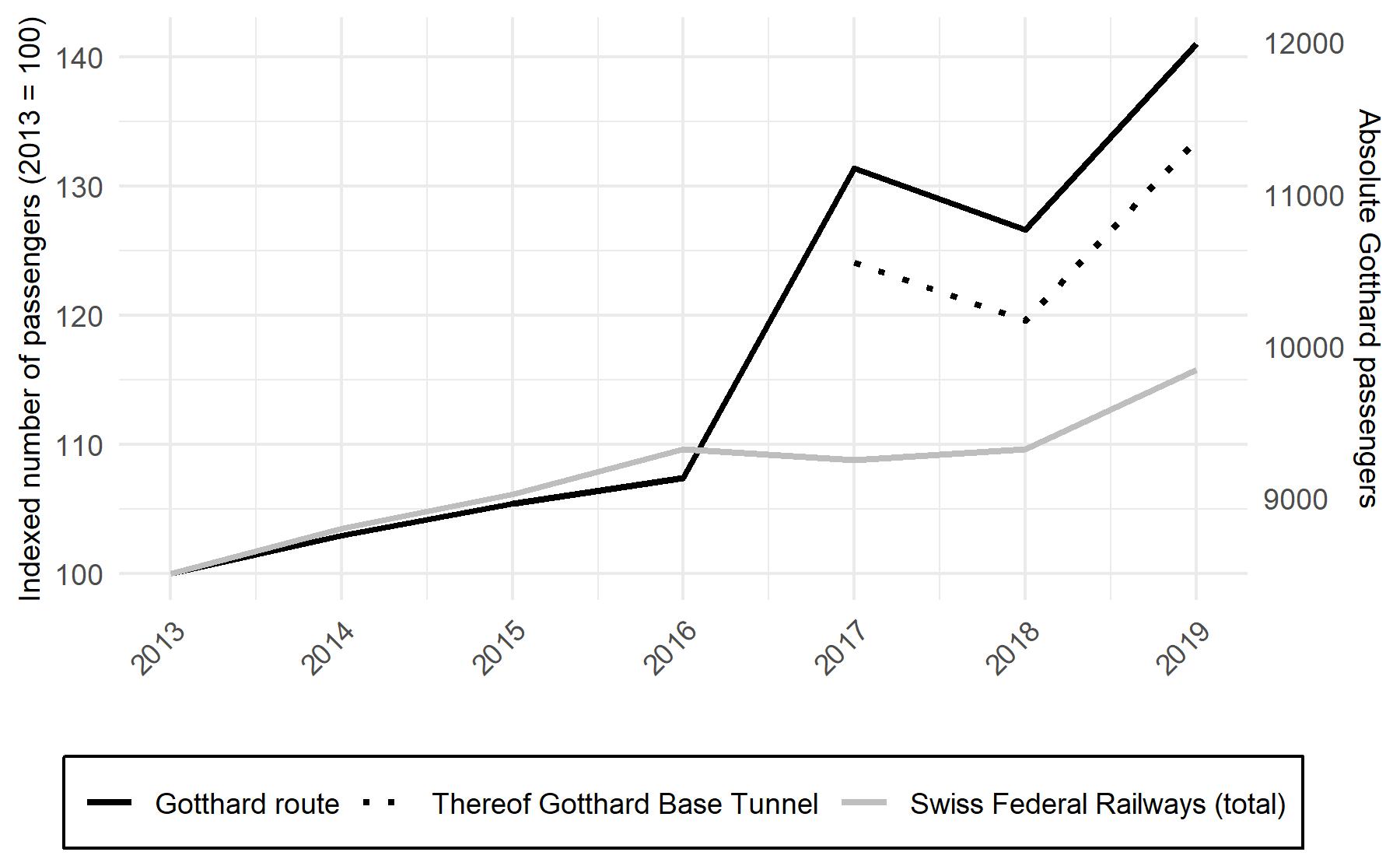}
\end{figure}

The modal split in passenger transport along the Gotthard axis after the commissioning of the Gotthard Base Tunnel shifted towards public transport usage \citep[for a descriptive analysis see][]{Monitoring2023}. The demand for public transportation increased by about 3,500 passengers per day from 2013 to 2019, rising from about 8,500 to 12,000, an increase of 41\%. From 2016 to 2017, the daily number of passengers increased from about 9,100 to 11,200, an increase of around 22\%. Figure \ref{Passengers_rail} shows that after 2016---i.e., following the opening of the Gotthard Base Tunnel---the number of rail passengers on the Gotthard route increased more rapidly than the average number of passengers on long-distance routes of the Swiss Federal Railways. By 2019, passenger numbers on the Gotthard route had reached 141\% of their 2013 level, whereas the national average had increased by only about 16\% over the same period. Moreover, in Figure \ref{Passengers_rail}, we also see that only about 5\% of the passengers used 2016 the Gotthard mountain route, whereas 95\% traveled through the Gotthard Base Tunnel.

\begin{figure}[H]
	\centering \caption{\label{Passengers_Goeschenen} Average Daily Road Traffic at the Gotthard Road Tunnel (Göschenen), April–October}\includegraphics[scale=.25]{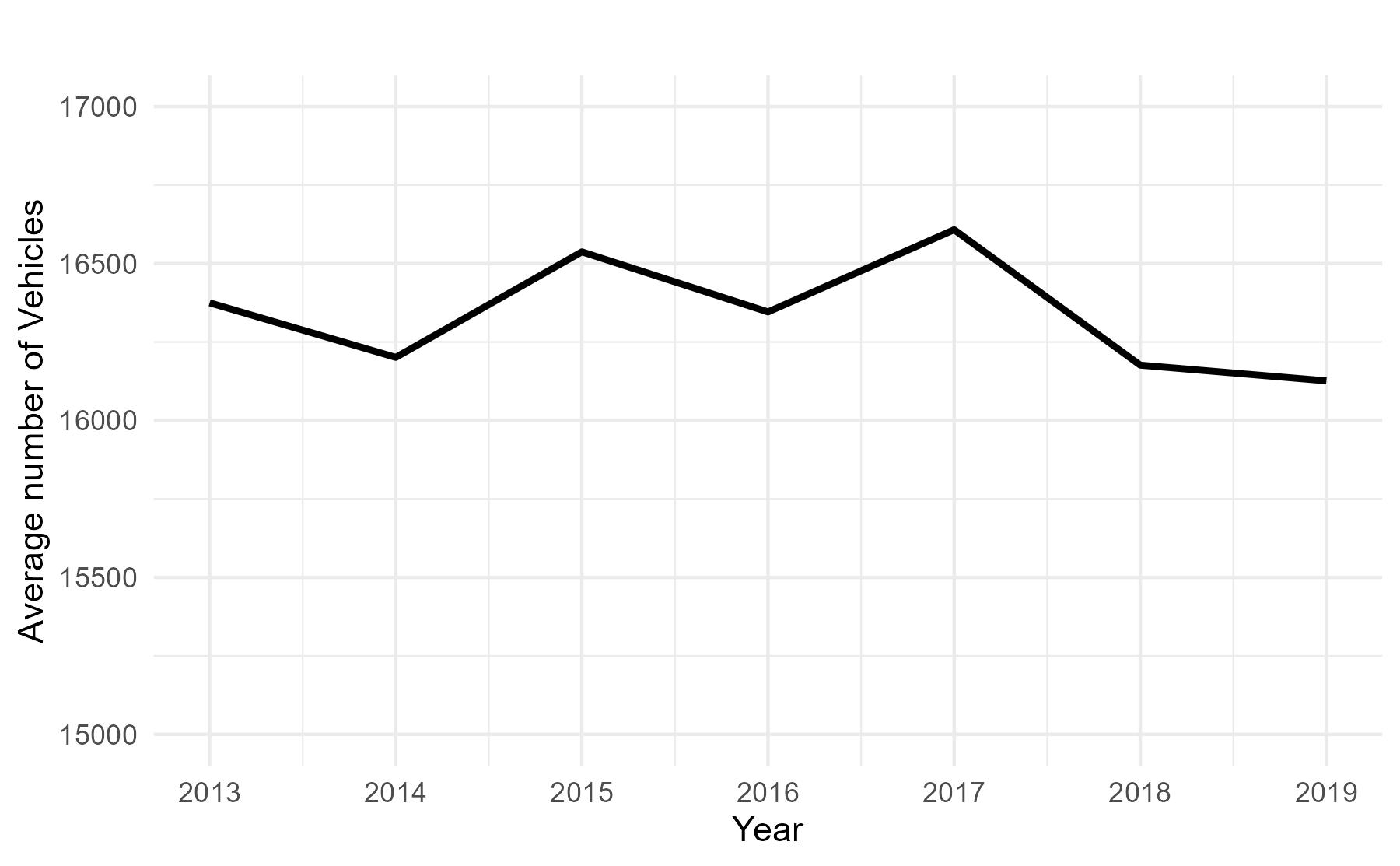}
\end{figure}

On the other hand, the demand for the parallel motorway section remained relatively stable during this period. Figure \ref{Passengers_Goeschenen} displays the average daily number of vehicles at the entrance of the Gotthard Road Tunnel in Göschenen for the months of April, May, June, July, September, and October. The number of vehicles fluctuates between approximately 16,000 and 16,700. In 2021, the average car occupancy in Switzerland was 1.53 persons per vehicle, with a higher rate observed in leisure travel, where the average occupancy reached 1.89 persons \citep{bfs_are2023}.

\section{Identification and estimation}\label{Methods}

\subsection{Methodology and implementation}\label{implement}

We aim to assess the effect of the Gotthard Base Tunnel on the parallel motorway section, i.e., the number of vehicles in the Gotthard road tunnel, which represents a single treated unit. While we can observe the vehicles in the Gotthard road tunnel, identifying the impact of the Gotthard Base Tunnel requires a comparison with a counterfactual---a scenario of the Gotthard motorway section to show what would have happened if the Gotthard Base Tunnel had not occurred. To construct this scenario, we use the synthetic control method \citep[see, e.g.,][]{abadie2010synthetic, abadie2021using}, selecting control units from other alpine crossings that did not receive comparable major rail transport infrastructure investments. These alpine crossings form a so-called donor pool, allowing us to create a 'synthetic Gotthard motorway section' that represents what might have occurred in on the Gotthard motorway section without the Gotthard Base Tunnel. This synthetic version is derived by assigning weights to each alpine crossing in the donor pool based on their similarity to the Gotthard alpine crossing prior to the Gotthard Base Tunnel's opening, aiming to minimize pre-treatment differences.

In this study, we employ a panel data set consisting of \(n\) alpine crossings, indexed by \(i \in \{ 1,...,n \}\), observed over \(T\) time periods, each denoted by \(t \in \{1,...,T\}\). The variable \(Y_{it}\) represents the observed outcome (i.e., number of vehicles) for unit \(i\) in period \(t\). The unit \(i=n\) (the Gotthard motorway section) experiences the effect of the new rail transport infrastructure starting in the post-treatment period \(T_0+1\), with \(T_0\) denoting the last pre-treatment period, i.e., 2016. According to \cite{huber2023causal}, we calculate the estimated effect of the intervention, \(\hat{\Delta}_{n,T=t}\), for the Gotthard in the periods following the intervention (\(t \geq T_0+1\)). This estimate is computed as the difference between the observed outcome on the Gotthard motorway section and a weighted average of outcomes from the non-intervened units---other alpine crossings---during the same periods. The synthetic control estimator \(\hat{\Delta}_{n,T=t}\) is thus defined as:

\begin{equation}
	\hat{\Delta}_{n,T=t}= Y_{nt}-\sum_{i=1}^{n-1} \hat{w}_iY_{it}, \text{for any } t \geq T_0+1. 
\end{equation}

We use the synthetic control method to derive the weights \(\hat{w}_i\) for each non-treated unit in a manner that the weighted average of their pre-intervention outcomes closely replicates the pre-intervention outcome of the Gotthard motorway section. The aim is to estimate what would have happened on the Gotthard motorway section in the absence of the Gotthard Base Tunnel. Therefore, the condition for the pre-treatment fit is:

\begin{equation}
	\sum_{i=1}^{n-1} \hat{w}_iY_{it} \approx Y_{nt}, \text{for all } t=1,...,T_0.
\end{equation}

The synthetic control method selects a weight vector  $W = (\hat{w}_1, \dots, \hat{w}_{n-1})$, with $\hat{w}_i \geq 0$ and $\sum_{i=1}^{n-1} \hat{w}_i = 1$. That means that the sum of all weights for the non-treated units sum up to one and are non-negative. The goal is to minimize the squared difference between the treated unit’s outcome 
and the weighted combination of control units over the pre-treatment period.

In addition to the synthetic control method, we also employ more recent synthetic difference-in-differences (SDID) approach as suggested by \citet{arkhangelsky2021synthetic}. 

To implement the synthetic control method using the statistical software \textsf{R}, we utilize the functions available in the \textit{synth} package as detailed by \cite{hainmueller2015synthPackage}, allowing to apply the synthetic control group method for comparative case studies. as described by \citet{abadie2003economic} and \citet{abadie2010synthetic}. Additionally, we apply the synthetic difference-in-differences method through the \textit{synthdid} package, provided by \cite{Arkhagelsky2021synthdid}, which allows the implementation of the SDID estimator for the average treatment effect in panel data, as proposed by \citet{arkhangelsky2019synthetic}.

\subsection{Assumptions}\label{Section:Assumptions}

Ensuring that our estimates identify the mechanism of interest, i.e., the effect of the Gotthard Base tunnel on the vehicles on the parallel motorway section, relies on assumptions about the data-generating process (see, e.g., \cite{huntington2021effect}). In the following, we present the identifying assumptions underlying our analysis (see also, e.g., \citet{abadie2021using} or more recently in the transport literature \citet{wallimann2024austria}):\newline
\textbf{Assumption 1 (no anticipation):} \newline
Assumption 1 is satisfied when the vehicles crossing the Gotthard motorway section did not change due to forward-looking individuals prior to the Gotthard Base tunnel. According to Assumption 1, tourists also must not alter their behavior prior to the completion of the Ceneri Base Tunnel (in the pre-treatment and post-treatment periods). \newline
\textbf{Assumption 2 (availability of a comparison group):} \newline
According to Assumption 2, the alpine crossings in the donor pool are sufficiently similar to the Gotthard. We discuss Assumption 2 in greater detail in Section \ref{Data} and Section \ref{results:sensitivity}.\newline
\textbf{Assumption 3 (convex hull condition):} \newline
By Assumption 3, the pre-treatment economic measures of the Gotthard road tunnel must not be too extreme compared to the other alpine crossings. The vehicles in the Gotthard road tunnel must not be (much) lower than the lowest and (much) higher than the highest outcome of the donor pool during the pre-treatment period. We discuss Assumption 3 in greater detail in Section \ref{Data}.\newline
\textbf{Assumption 4 (no spillover effects):} \newline
Assumption 4 implies that there must be no spillover effects of the Gotthard Base tunnel on other alpine crossings. However, it is conceivable that, for example, a traveler journeying from Hamburg to Milan might alter her route if less congestion is expected on the Gotthard route. For instance, on April 30, 2025, at 11 p.m., travel via the Gotthard takes according to Google Maps 11 hours and 57 minutes, while using the San Bernardino Tunnel takes 11 hours and 48 inutes—making the two routes nearly identical in terms of travel time. Nevertheless, during peak hours, the Gotthard motorway section may now be somewhat less congested, meaning that travelers, after the tunnel’s introduction, may opt not to switch to the San Bernardino Tunnel. Similar spillover effects might occur for other journeys. Therefore, we assess the sensitivity of our results regarding Assumption 4 in greater detail in Section \ref{results:sensitivity}. \newline
\textbf{Assumption 5 (no external shocks):} \newline
Assumption 5 is satisfied when no external regional shocks occur to the outcome of interest, i.e., vehicles (in one or only several treatment or control units), during the study period.

\section{Data}\label{Data}

The data set comprises monthly counts of vehicle crossings at major Alpine transit points from April to October, covering 2013 to 2019. The focus on these months reflects the tourist season for travel to Ticino and Italy. The focus on the period from April to October is due to the limited usability of several Alpine crossings during the winter months. Seasonal road closures affecting Alpine crossings in the control group could introduce bias if comparisons were made over the entire year. The perimeter of interest is defined by the Gotthard motorway section beginning at the entrance of the Gotthard Road Tunnel in Göschenen and ending at its exit in Airolo. The number of vehicles on this section is measured by the automatic counting station 150 in Göschenen, which records traffic on the Gotthard route (i.e., the treatment group) in both directions. 

The Alpine crossings included in the donor pool are: Bernina (Switzerland), Brenner (Austria–Italy), Flüela (Switzerland), Fréjus Tunnel (France–Italy), Grand-Saint-Bernard (Switzerland–Italy), Julier (Switzerland), Karawanken Tunnel (Austria–Slovenia), Mont-Blanc Tunnel (France–Italy), San Bernardino Tunnel (Switzerland), and Tauern Tunnel (Austria).\footnote{For the data sources see \href{https://www.astra.admin.ch/astra/de/home/dokumentation/daten-informationsprodukte/verkehrsdaten/daten-publikationen/automatische-strassenverkehrszaehlung/monats-jahresergebnisse.html}{https://www.astra.admin.ch/astra/de/home/dokumentation/daten-informationsprodukte/verkehrsdaten/daten-publikationen/automatische-strassenverkehrszaehlung/monats-jahresergebnisse.html}, \href{https://tunnelmb.net/public/files/430/5-5-download-1-traffico-mensile-2018-2019.pdf}{https://tunnelmb.net/public/files/430/5-5-download-1-traffico-mensile-2018-2019.pdf}, \href{https://www.sitaf.it/dati-di-traffico-t4/}{https://www.sitaf.it/dati-di-traffico-t4/}, and \href{https://www.asfinag.at/verkehr-sicherheit/verkehrszaehlung/}{https://www.asfinag.at/verkehr-sicherheit/verkehrszaehlung/}, accessed on April 29, 2025.} Data for years prior to 2013 contain missing values, and years after 2019 are excluded due to distortions caused by the COVID-19 pandemic. Therefore, the analysis period spans from April 2013 to October 2019. Note that we focus on monthly data, which report the average daily traffic of light vehicles weighing less than or equal to 3.5 tonnes. 

In Figure \ref{Average_resid} in Appendix \ref{Appendix_Figures}, we see that the Tauern Tunnel (Austria), Brenner (Austria–Italy), and Karawanken Tunnel (Austria–Slovenia) exhibit a distinctly different, specifically, more positive development compared to the other Alpine crossings. Therefore, as Assumption 2 requires alpine crossings in the donor pool that are sufficiently similar to the Gotthard motorway section, we exclude these from the analysis.

If we look at Panel A in Table \ref{tab:residuals_vehiclecounts}, we see that the pre-treatment outcome of the Gotthard road tunnel is with an average of 16,365 vehicles per day, distinctly higher than the vehicles on other alpine crossings (mean: 3333 vehicles). Also, when considering the maximum of vehicles in the control group, we see that the Gotthard is more extreme in the values of the outcome variable, the number of vehicles. Therefore, when considering vehicles, Assumption 3 is violated, and we are not able to construct a weighted average of untreated units that can approximate the outcome variable for the treated unit before the treatment \citep[see, e.g., ][]{abadie2021using}. 

However, there exist other ways to proceed in such cases \citep[see some examples discussed by ][]{abadie2021using}. In our study, we proceed by computing monthly residuals for each Alpine crossing \( i \). Therefore, we subtract pre-treatment monthly averages from the observed traffic values. Let \( t \) index the monthly observations, with \( y(t) \) and \( m(t) \) denoting the corresponding year and month, respectively. We define the residual as:

\[
\text{resid}_{it} = \text{vehicle}_{it} - \bar{v}_{i,m(t)}
\]

where \( \bar{v}_{i,m(t)} \) is the average traffic in month \( m \) at crossing \( i \), computed using only data from years prior to 2017, i.e., the pre-treatment period. These residuals are then averaged by year and crossing:

\[
\overline{\text{resid}}_{iy} = \frac{1}{N_{iy}} \sum_{t \in T_{iy}} \text{resid}_{it}
\]

where \( T_{iy} = \{ t : y(t) = y \} \) is the set of all months observed for crossing \( i \) in year \( y \). The resulting series \( \overline{\text{resid}}_{iy} \) captures year-specific deviations from typical seasonal patterns, allowing for comparisons across years and Alpine crossings.

\begin{table}[H]
	\centering
	\caption{Descriptive statistics of residuals and vehicle counts per street, by group and period (April-October)} \label{tab:residuals_vehiclecounts}
	\begin{tabular}{llccccc}
		\toprule
		\textbf{Group} & \textbf{Period} & \textbf{Mean} & \textbf{SD} & \textbf{Min} & \textbf{Max} & \textbf{N} \\
		\midrule
		\multicolumn{7}{l}{\textit{Panel A: Mean actual vehicle counts}} \\
		Gotthard   & Pre-treatment   & 16365  & 138.0 & 16201 & 16537 & 4 \\
		Gotthard   & Post-treatment  & 16303  & 265.0 & 16126 & 16608 & 3 \\
		Donor Pool & Pre-treatment   & 3333   & 1822.0 & 1845  & 7699  & 28 \\
		Donor Pool & Post-treatment  & 3505   & 1847.0 & 1841  & 8032  & 21 \\
		\multicolumn{7}{l}{\textit{Panel B: Mean residuals of vehicle counts}} \\
		Gotthard   & Pre-treatment   & 0.00   & 138.0 & -164 & 173  & 4 \\
		Gotthard   & Post-treatment  & -61.3  & 265.0 & -238 & 243  & 3 \\
		Donor Pool & Pre-treatment   & 2.61   & 144.0 & -292 & 347  & 28 \\
		Donor Pool & Post-treatment  & 113.0  & 181.0 & -363 & 516  & 21 \\
		\bottomrule
	\end{tabular}
\end{table}

In Table \ref{tab:residuals_vehiclecounts}, we now observe that the average number of daily vehicles on the Gotthard route decreases in the post-treatment period, from 0 vehicles to -61.3 vehicles. In contrast, the donor pool units show an increase, from 2.61 vehicles in the pre-treatment period to 113 vehicles post-treatment. Regarding Assumption 3, and when looking at minimum and maximum number of vehicles during the pre-treatment period, we see that the Gotthard road tunnel is now not lower than the lowest and higher than the hightest outcome of the donor pool. 

\section{Results}\label{Results}

\subsection{Main Results}

\begin{figure}[H]
	\centering \caption{\label{SCM_year} Demand development of the Gotthard motorway section and the synthetic counterpart}\includegraphics[scale=.5]{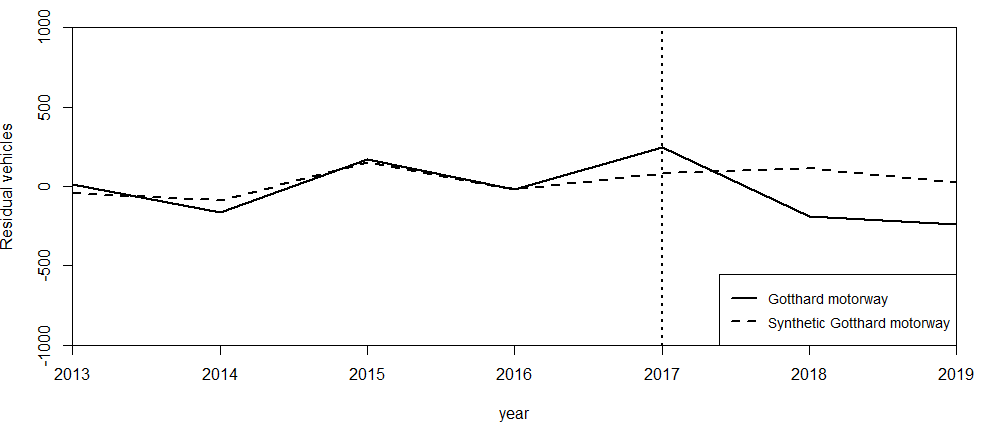}
\end{figure}

Figure \ref{SCM_year} shows the main result when applying the synthetic control method. The trajectories of the Gotthard motorway section and the synthetic Gotthard motorway section mainly track each other closely during the period preceding the opening of the Gotthard Base Tunnel at the end of 2016 (and, therefore, including 2016). Therefore, we are able to mimic the counterfactual outcome of how the vehicles on the Gotthard motorway section would have evolved in the absence of the Gotthard Base tunnel. The synthetic control method assigns a non-zero weight of 76.8\% to the Bernina and a weight of 23.2\% to the Fréjus Tunnel. All other alpine crossings get a weight of 0\%.

Moreover, Figure \ref{SCM_year} also indicates that the number of vehicles (i.e., the average residuals compared to the pre-treatment period) on the Gotthard motorway section were lower compared to the synthetic Gotthard motorway section. On average, the daily number of vehicles was 135 vehicles lower during the months April to October. In other words, the traffic decreased by 0.8\% compared to the pre-treatment average of 16,365 daily vehicles (see Table \ref{tab:residuals_vehiclecounts}). 

To determine the statistical significance of the treatment effect, we randomly draw seven unists with replacement from the donor pool 200 times to arrive at bootstrap confidence intervals \citep[see, e.g., ][for an example in the transportation literature]{wallimann2023price}. The corresponding 95\% bootstrap confidence interval of the average estimated effect is [-269; -127] and, thus, does not include zero. Therefore, we conclude that the effect is statistically significantly different from zero at the conventional 5\% level.

When applying the synthetic difference in differences method, we achieve at a comparable negative effect amounting to a decrease of 152 vehicles per day. Here, the corresponding 95\% bootstrap confidence interval of the average estimated effect is [-291; -84].

\subsection{Sensitivity analysis}\label{results:sensitivity}

Assumption 4 states that there are no spillover effects from the Gotthard Base Tunnel on other alpine crossings. To address a potential spillover effect as discussed in Section \ref{Section:Assumptions}, we exclude the San Bernardino route in our \textit{first} sensitivity analysis. 

The effect of the new Gotthard Base Tunnel on car traffic, when applying the synthetic control method, amounts to a reduction of 135 vehicles per day, or 0.8\% compared to the pre-treatment mean of vehicles, which is the same as in the original result, as the San Bernardino got a zero weight. The 95\% bootstrap confidence interval now amounts to [–269, -110] and is again statistically significant at the conventional 5\% level. 

Note that when using the synthetic control method to construct our Synthetic Gotthard motorway section, we include the pre-treatment values of the dependent variable for all years in the pre-treatment period from 2013 to 2016. This allows the method to match the entire pre-treatment trajectory, rather than just the average. Therefore, as a \textit{second} sensitivity analysis, we exclude the pre-treatment predictors and instead match solely on the outcome variable over the full pre-treatment period.\footnote{This corresponds to deactivating the \texttt{special.predictors} argument in the \texttt{dataprep} command in the statistical software \textsf{R}.} The estimated effect of the new Gotthard Base Tunnel on traffic amounts now to \(-152\) vehicles per day---or approximately 1\% relative to pre-treatment traffic levels---with a 95\% bootstrap confidence interval of [\(-265\); \(-89\)].

According to Assumption 2, we assume that the units in the donor pool are sufficiently similar to the Gotthard motorway section. To examine whether the introduction of the Gotthard Base Tunnel had an effect on domestic leisure traffic by car, we conducted a \textit{third} sensitivity analysis by assembling a different donor pool. For this purpose, we selected touristic routes in Switzerland.\footnote{Unfortunately, for many routes, data are not available for the full period.} Considering yearly averages, the data set contains the Gotthard motorway section and selected counting stations Alpnachstad, Erstfeld, Flüela, Gondo, Gstaad, Hinterrhein, Schattdorf, and Tamis. 

\begin{figure}[H]
	\centering \caption{\label{SCM_year_touri} Demand development of the Gotthard motorway section and the synthetic counterpart using touristic routes from Switzerland as donor pool}\includegraphics[scale=.5]{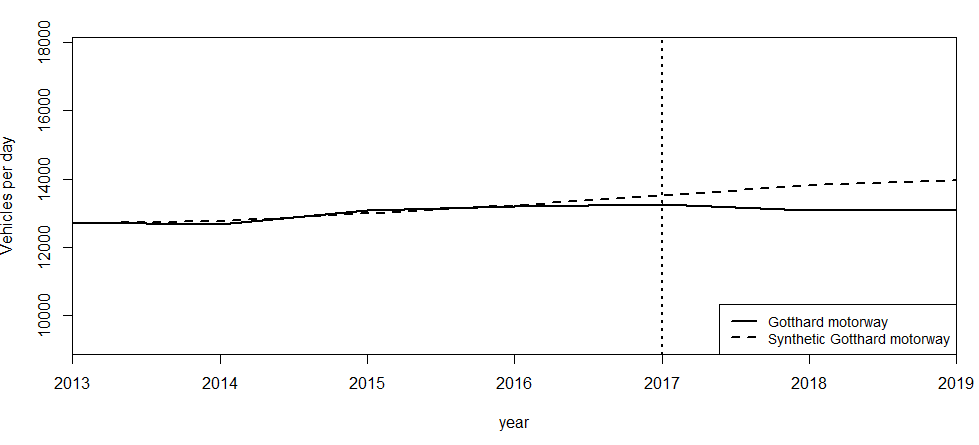}
\end{figure}

In Figure \ref{SCM_year_touri} we see that the trajectories of the Gotthard and synthetic Gotthard motorway sections again align closely prior to the intervention. We observe a negative effect of –617 daily vehicles attributable to the Gotthard Base Tunnel. This effect is now noticeably larger negative, corresponding to approximately a 5\% reduction in traffic compared to the 2016 average of 13,200 daily vehicles. However, the 95\% bootstrap confidence interval for the estimated effect of [-785; 2008] includes zero and the effect should therefore be interpreted with caution.

\section{Discussion and Conclusion}\label{Discussion}

In this study, we employed the synthetic control and synthetic difference-in-differences methods to estimate the causal impact of the Gotthard Base Tunnel on the parallel Gotthard motorway section. By constructing a counterfactual based on a weighted combination of alternative alpine crossings in Switzerland and neighboring countries, we estimated that the number of light vehicles decreased by 135 and 152 per day, respectively, during the months from April to October following the tunnel’s opening. Considering the average leisure car occupancy rate of 1.89, this corresponds to approximately 270 individuals per day. These figures represented an a decline in vehicles of just under 1\% compared to the pre-intervention baseline. Although the magnitude of the effect was modest, bootstrap confidence intervals indicated statistical significance. Further sensitivity analysis, e.g., accounting for potential spillover effects reinforced these findings. Additionally, when constructing a counterfactual with other Swiss motorway sections with similar exposure to leisure traffic---instead of using other alpine crossings in Switzerland and neighboring countries---, the estimated reduction in daily vehicles rose to 617. However, it is no longer statistically significant. In summary, in contrast to the study of \citet{borsati2020modal} analyzing the case of high-speed rails (HSR) in Italy, the construction of an efficient railway infrastructure, i.e., the Gotthard Base Tunnel with its travel time savings had a modal shift effect.

However, between 2013 and 2019, the number of rail passengers on the Gotthard route increased by about 41\%. This growth in demand also exceeds the average increase observed across the Swiss Federal Railways network. Comparing this number with the effect on the parallel Gotthard motorway section, we see that the decrease on the Gotthard motorway section is substantially smaller than the increase observed on the rail network. In conclusion, our findings suggest that, on such touristic routes, road and rail transport are not perfect substitutes. Put differently, the substitution from the car is modest, which is in line with the literature \citep{givoni2013review}. 

\begin{figure}[H]
	\centering \caption{\label{ModeShare} Decomposition of the 2019 public transport share}\includegraphics[scale=.35]{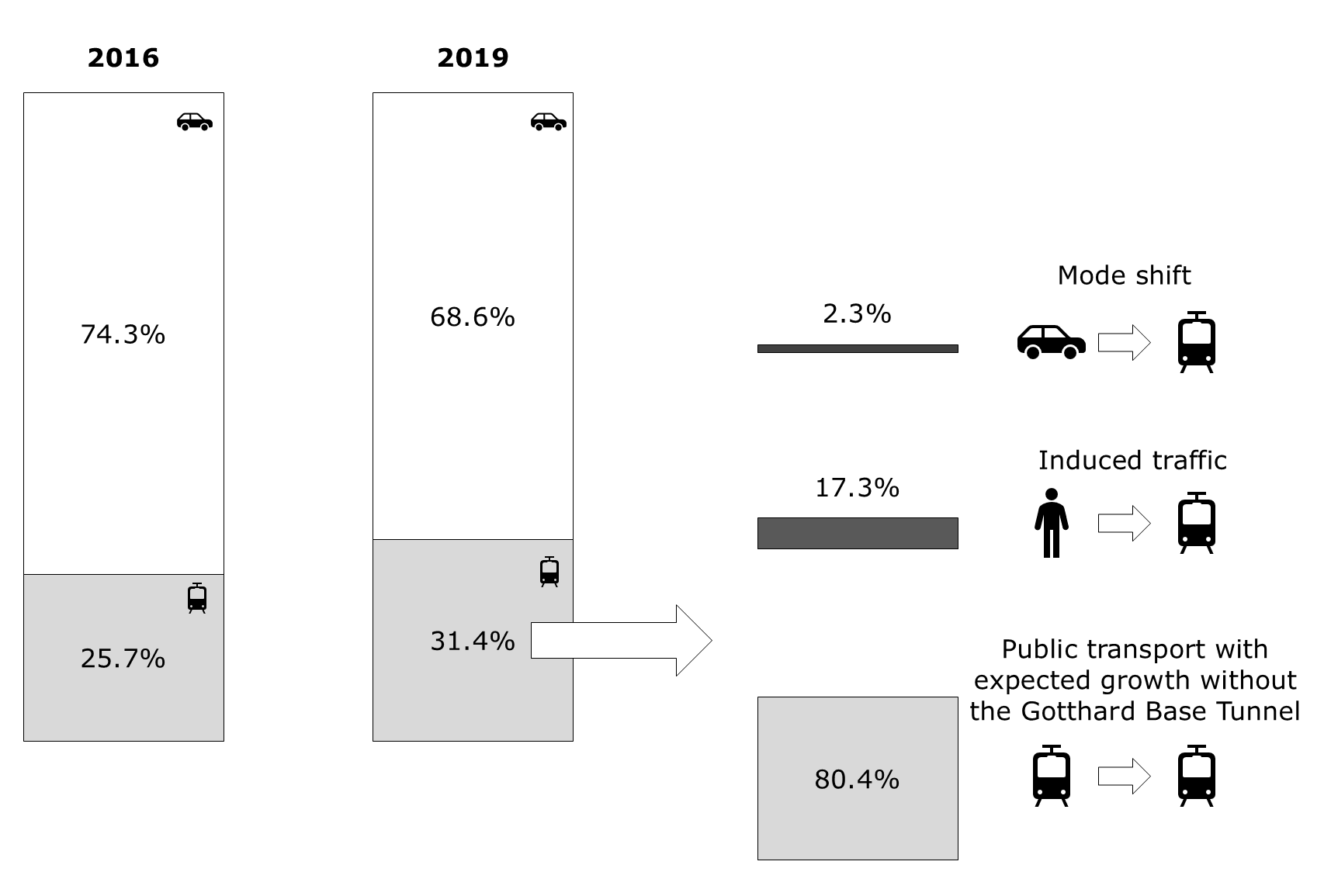}
\end{figure}

However, with the Gotthard Base Tunnel, only pull measures were implemented. When looking at mode share on the Gotthard route, we observe a rise in the public transport share from 25.7\% in 2016 to 31.4\% in 2019 (see Figure \ref{ModeShare}). However, only approximately 270 individuals shifted from car to public transport per day, corresponding to just 2.3\% of all public transport users. Moreover, assuming identical growth rates as in the Swiss long-distance public transport sector, 80.4\% of users would have traveled by public transport anyway. This leaves 17.3\% as induced demand, i.e., new public transport users who would not have taken the Gotthard route in the absence of the Gotthard Base Tunnel. In conclusion, achieving a substantial modal shift through pull measures alone is difficult; they must be complemented by push measures such as road pricing.

Looking at Table \ref{tab:travel_time_gbt} in Appendix \ref{Appendix_Tables}, the train remains slower than the car on several routes. Therefore, if the dataset could be restricted to city trips only (or at least to trips originating in urban areas), the estimated effect would likely be relatively larger. Moreover, tourists are often traveling in groups. In such cases, traveling by car is often more cost-effective, especially when none of the passengers hold public transport subscription (e.g., a Half Fare Travelcard). Additionally, with regard to transit travel, there are no direct train connections between Germany and Italy, or vice versa. Therefore, our result is not completely surprising \citep{chang2008accessibility}. Also that tourists switch to rail during periods of high north-south (tourist) traffic in order to avoid congestion---thereby reducing peak demand at the road tunnel---cannot be confirmed based on the diurnal traffic patterns shown in Figure \ref{Tagesgang_2016} and Figure \ref{Tagesgang_2019} in Appendix \ref{Appendix_Figures}.

The short-term perspective, spanning three years, outlined in this paper constitutes the main limitation of our study. It would be of particular interest to examine the long-term effects (e.g., more than ten years), as well as to assess whether the negative impact on the number of vehicles on the parallel Gotthard motorway section intensified following the completion of the Ceneri Base Tunnel, which further increased train frequency and reduced travel times---developments that are also especially relevant for transit traffic, a key segment of travel along the Gotthard corridor. However, this extension of our study was not feasible due to the external shock of the COVID-19 pandemic. Future empirical research on similar natural experiments should aim to investigate the long-term effects of public transport quality improvements on road traffic.

Another limitation of our study is that we do not account for treatment effect heterogeneity. It is possible that different groups respond differently to the intervention. For instance, \citet{blattler2024free} show that domestic tourists in Switzerland who regularly travel multi-modal in their daily lives are more responsive to incentives---such as free public transport for arrival and departure---and are therefore more likely to switch from car to public transport. It is conceivable that such individuals were also more responsive to the reduction in travel time, whereas those with little connection to public transport in their everyday life did not react at all.

	\newpage
	\bigskip
	
	\bibliographystyle{econometrica}
	\bibliography{GBT.bib}
	
	\bigskip
	\newpage
	
\begin{appendix}
		
		\numberwithin{equation}{section}
		\counterwithin{figure}{section}
		\noindent \textbf{\LARGE Appendices}
	
	\section{Additional Tables}\label{Appendix_Tables}
	
	\begin{table}[H]
		\centering
		\caption{Travel Time Comparison: Car vs. Train Before and After the Gotthard Base Tunnel (GBT)} \label{tab:travel_time_gbt}
		\small
		\begin{tabular}{llccccc}
			\toprule
			\textbf{Departure} & \textbf{Arrival} & \textbf{Car Time} & \textbf{Train 2016} & \textbf{Train 2019} & \textbf{Time Saved} & \textbf{Time Saved} \% \\
			\midrule
			Bern     & Bellinzona & 2:50 & 11:00--14:25 (3:25) & 09:00--11:57 (2:57) & 0:28 & 14\% \\
			Lucerne  & Bellinzona & 1:40 & 12:18--14:25 (2:07) & 10:18--11:57 (1:39) & 0:28 & 22\% \\
			Basel    & Bellinzona & 2:50 & 11:04--14:25 (3:21) & 09:04--11:57 (2:53) & 0:28 & 14\% \\
			Zurich   & Bellinzona & 2:10 & 12:09--14:25 (2:14) & 09:32--11:27 (1:55) & 0:19 & 14\% \\
			Bern     & Lugano     & 3:10 & 11:00--14:50 (3:50) & 09:00--12:26 (3:26) & 0:24 & 10\% \\
			Lucerne  & Lugano     & 2:00 & 12:18--14:50 (2:32) & 10:18--12:26 (2:08) & 0:24 & 16\% \\
			Basel    & Lugano     & 3:00 & 11:04--14:50 (3:46) & 09:04--12:26 (3:22) & 0:24 & 11\% \\
			Zurich   & Lugano     & 2:30 & 12:09--14:50 (2:39) & 09:32--11:56 (2:24) & 0:15 & 9\%  \\
			Bern     & Locarno    & 3:00 & 11:00--14:56 (3:56) & 09:00--12:29 (3:29) & 0:27 & 11\% \\
			Lucerne  & Locarno    & 1:50 & 12:18--14:56 (2:38) & 10:18--12:29 (2:11) & 0:27 & 17\% \\
			Basel    & Locarno    & 3:00 & 11:04--14:56 (3:52) & 09:04--12:29 (3:25) & 0:27 & 12\% \\
			Zurich   & Locarno    & 2:20 & 12:09--14:56 (2:45) & 09:32--11:59 (2:27) & 0:18 & 11\% \\
			Bern     & Milan      & 3:50 & 10:04--15:35 (5:31) & 11:00--15:50 (4:50) & 0:41 & 12\% \\
			Lucerne  & Milan      & 2:40 & 11:40--15:35 (3:55) & 12:18--15:50 (3:32) & 0:23 & 10\% \\
			Basel    & Milan      & 3:50 & 10:17--15:35 (5:18) & 11:04--15:50 (4:46) & 0:32 & 10\% \\
			\bottomrule
		\end{tabular}
		\begin{tablenotes}[flushleft]
			\footnotesize
			\item \textit{Note:} Train travel durations are calculated from departure to arrival times. Train connections were derived from the timetable archive available at \url{https://www.oev-info.ch/de/fahrplan-aktuell/fahrplanarchiv}. Time savings indicate reductions between 2016 and 2019 but may vary depending on the specific connection. For train journeys, the main station was used as the reference point; for car trips, the location specified by Google Maps was used. Car travel times refer to the shortest suggested driving time according to Google Maps for Monday, April 8, 2019, with a departure at 11:00 AM. Note that if we would consider access time to the train would make this difference even more pronounced.
		\end{tablenotes}
	\end{table}
	
	\section{Additional Figures}\label{Appendix_Figures}

	\begin{figure}[H]
		\centering \caption{\label{Average_resid} Residual Time Series per Alpine Route}	\includegraphics[scale=.25]{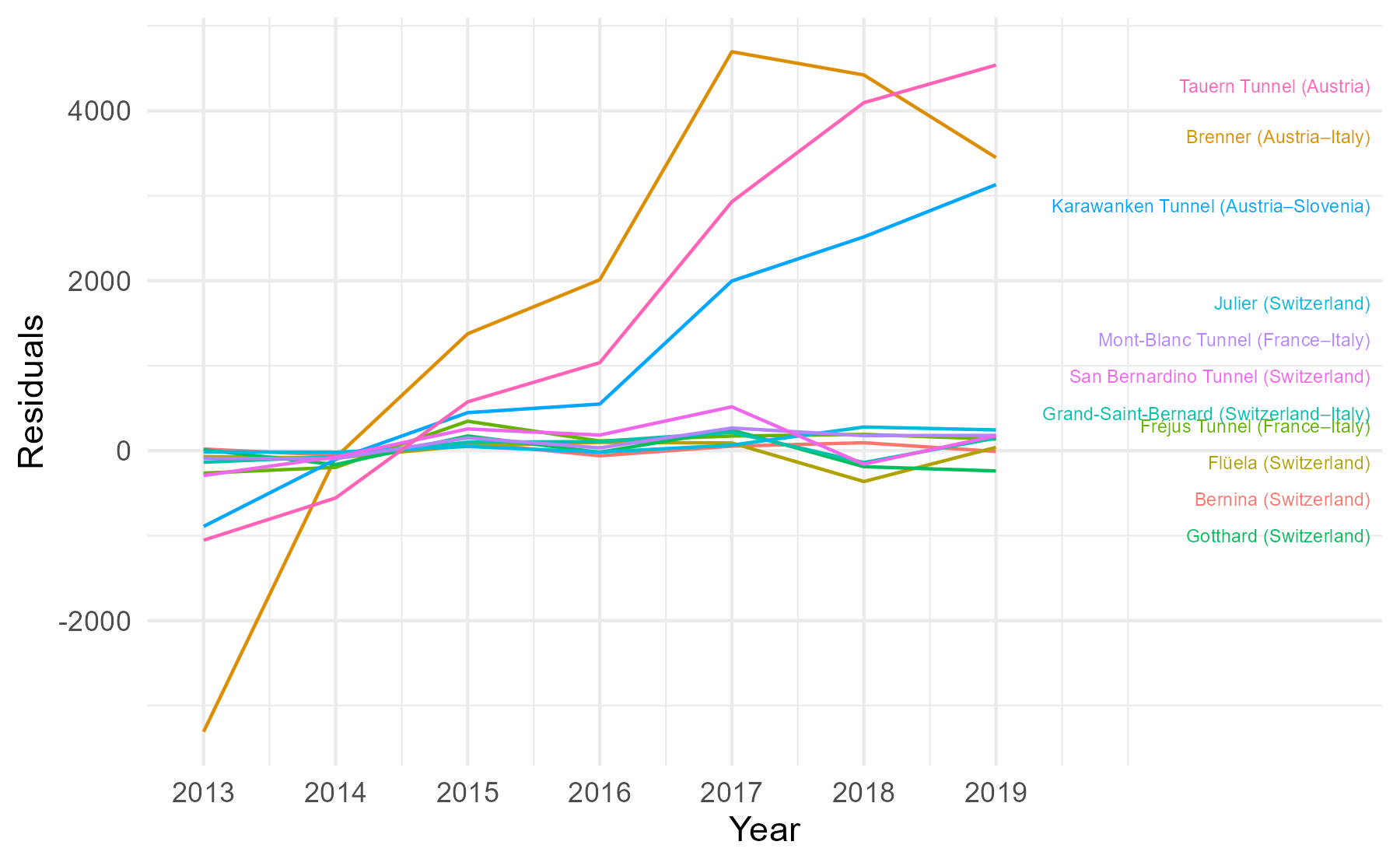}
	\end{figure}
	
	\begin{figure}[H]
		\centering \caption{\label{Tagesgang_2016} Average Hourly Vehicle Counts at Göschenen Counting Station by Day Type (Direction: Bellinzona, 2016)}	\includegraphics[scale=.25]{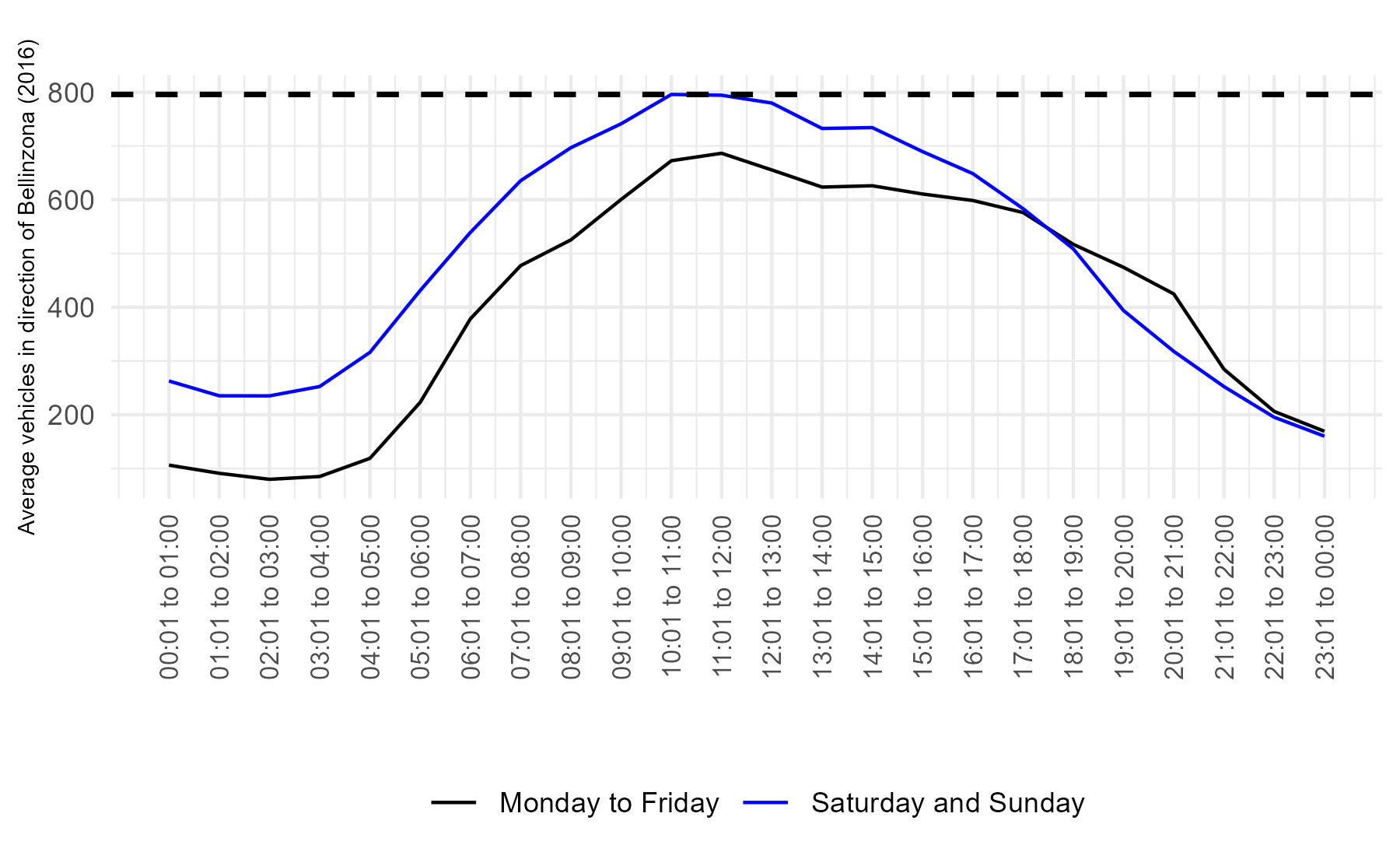}
	\end{figure}
	
	\begin{figure}[H]
		\centering \caption{\label{Tagesgang_2019} Average Hourly Vehicle Counts at Göschenen Counting Station by Day Type (Direction: Bellinzona, 2019)}	\includegraphics[scale=.25]{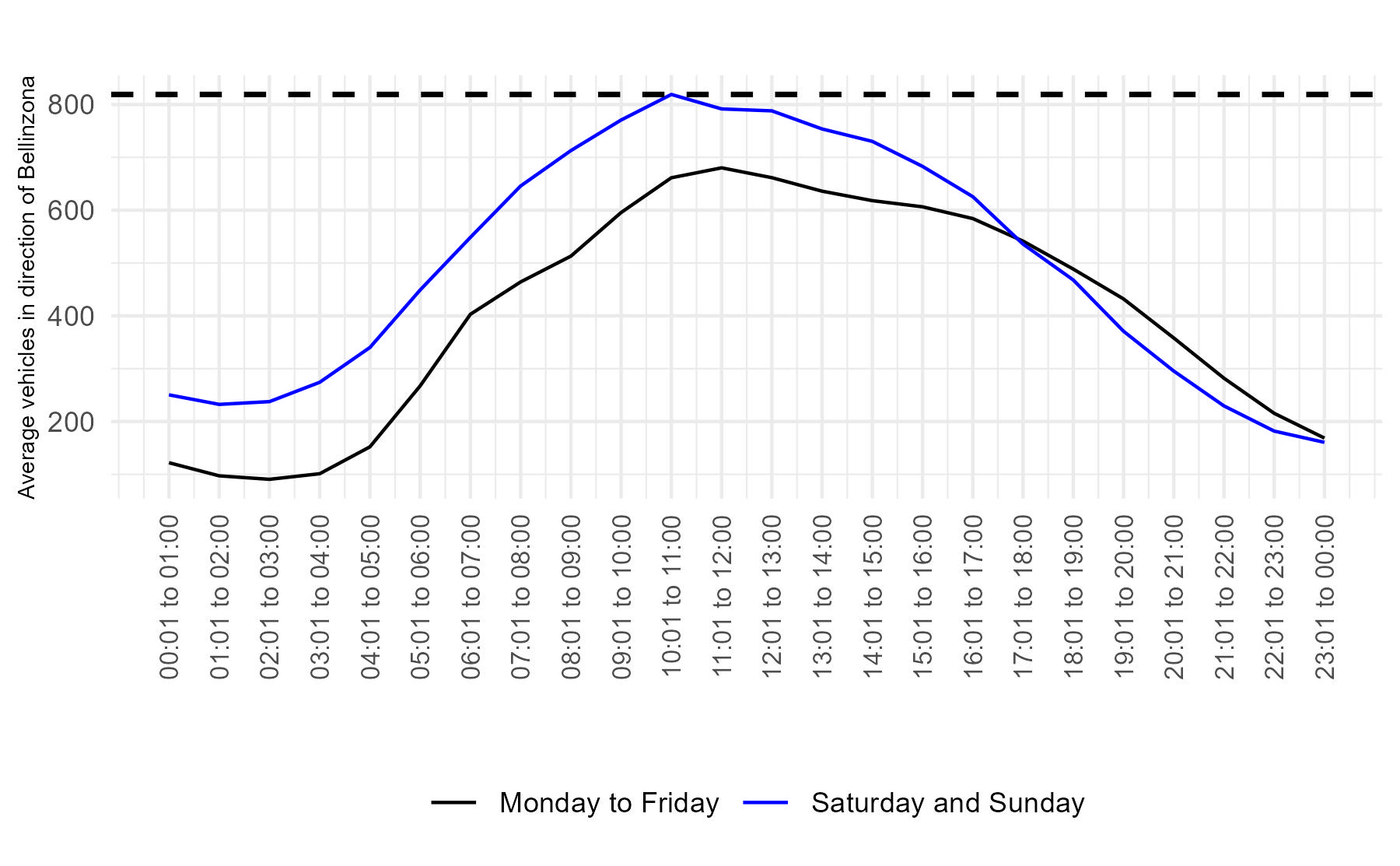}
	\end{figure}
	
		\begin{figure}[H]
		\centering \caption{\label{Boxplot_2016} Boxplot of Hourly Vehicle Counts at the Göschenen Counting Station, by Day Type (2016, Values < 2000)}	\includegraphics[scale=.25]{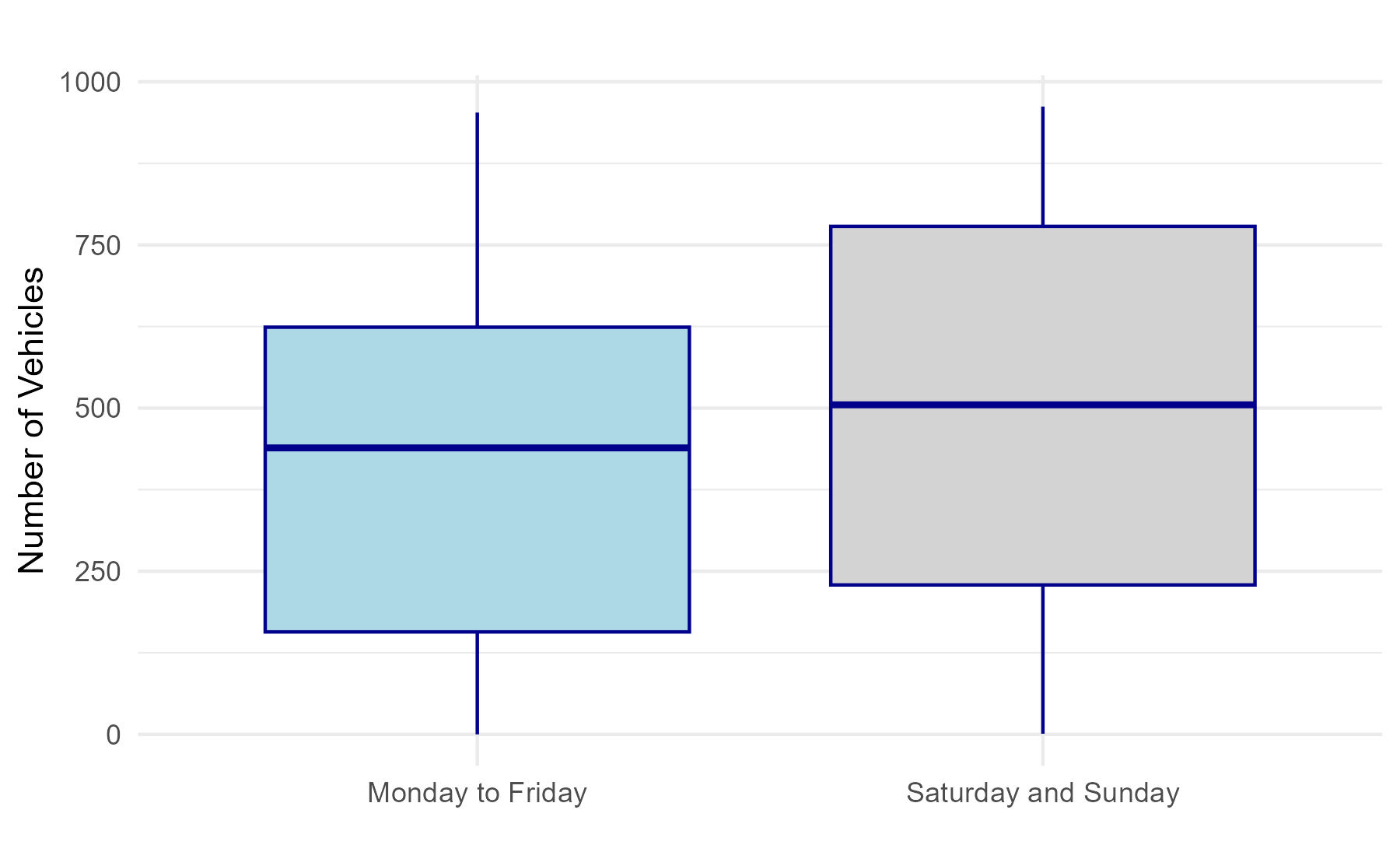}
	\end{figure}
	
	\begin{figure}[H]
		\centering \caption{\label{Boxplot_2019} Boxplot of Hourly Vehicle Counts at the Göschenen Counting Station, by Day Type (2019, Values < 2000)}	\includegraphics[scale=.25]{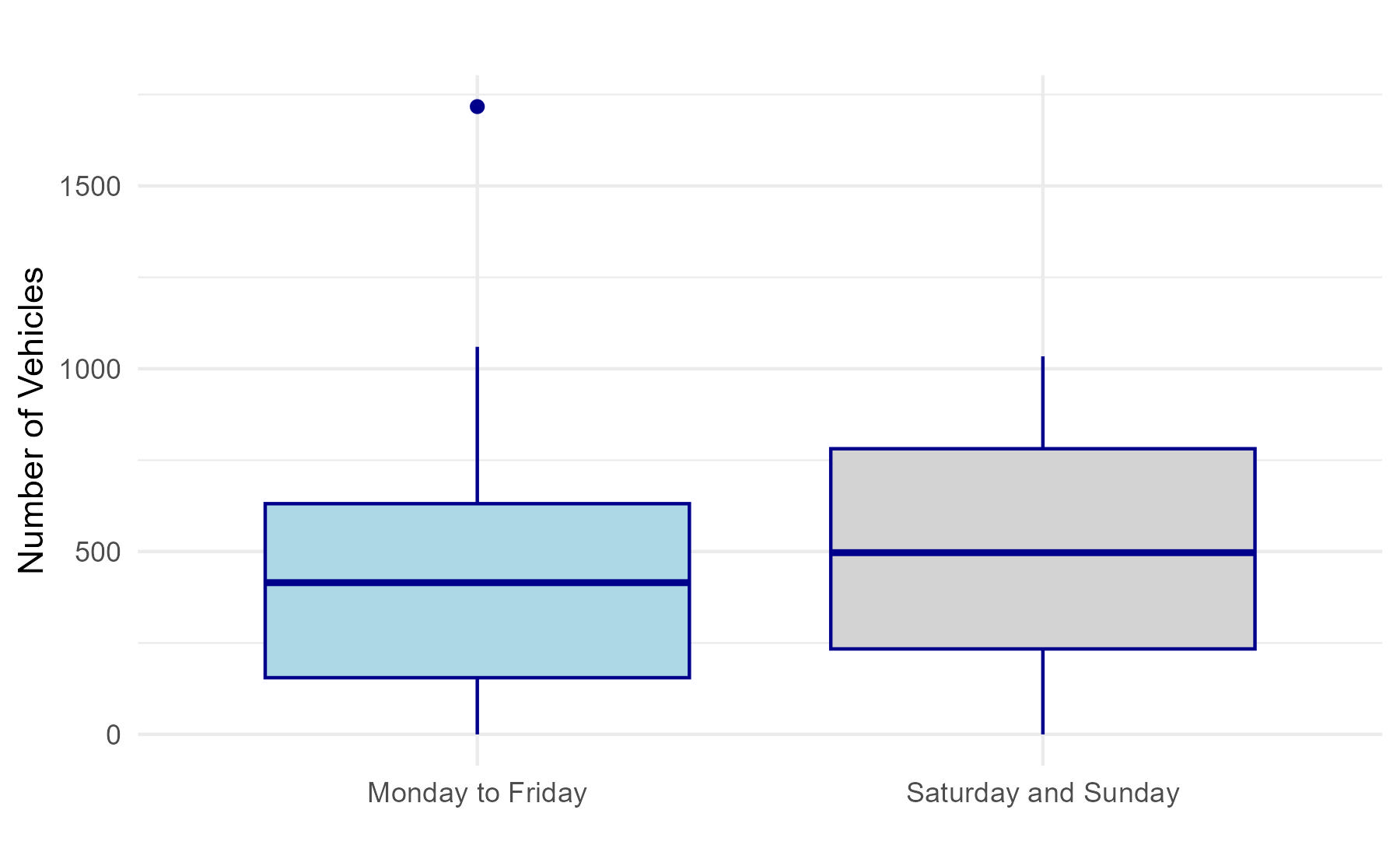}
	\end{figure}

	\end{appendix}
\end{document}